\documentclass[iicol, pdflatex, sn-mathphys-num]{sn-jnl}% Math and Physical Sciences Numbered Reference Style 
\usepackage{graphicx}%
\usepackage{multirow}%
\usepackage{amsmath,amssymb,amsfonts}%
\usepackage{amsthm}%
\usepackage{mathrsfs}%
\usepackage[title]{appendix}%
\usepackage{xcolor}%
\usepackage{textcomp}%
\usepackage{manyfoot}%
\usepackage{booktabs}%
\usepackage{algorithm}%
\usepackage{algorithmicx}%
\usepackage{algpseudocode}%
\usepackage{listings}%

\begin{document}
\title{Nucleation transitions in polycontextural networks towards consensus}
%%=============================================================%%
%% GivenName	-> \fnm{Joergen W.}
%% Particle	-> \spfx{van der} -> surname prefix
%% FamilyName	-> \sur{Ploeg}
%% Suffix	-> \sfx{IV}
%% \author*[1,2]{\fnm{Joergen W.} \spfx{van der} \sur{Ploeg} 
%%  \sfx{IV}}\email{iauthor@gmail.com}
%%=============================================================%%

\author*[1]{\fnm{Johannes} \sur{Falk} }\email{jfalk@constructor.university}
\author[2,3]{\fnm{Edwin} \sur{Eichler} }
\author[3,4]{\fnm{Katja} \sur{Windt} }
\author[1]{\fnm{Marc-Thorsten} \sur{Hütt}}
\affil[1]{School of Science, Constructor University, Bremen, Germany}
\affil[2]{EICHLER Consulting AG, Weggis, Switzerland}
\affil[3]{SMS Group GmbH, Düsseldorf, Germany}
\affil[4]{School of Business, Social and Decision Sciences, Constructor University, Bremen, Germany}

\abstract{Recently, we proposed polycontextural networks as a model of evolving systems of interacting beliefs. Here, we present an analysis of the phase transition as well as the scaling properties. The model contains interacting agents that strive for consensus, each with only subjective perception. Depending on a parameter that governs how responsive the agents are to changing their belief systems the model exhibits a phase transition that mediates between an active phase where the agents constantly change their beliefs and a frozen phase, where almost no changes appear. We observe the build-up of convention-aligned clusters only in the intermediate regime of diverging susceptibility. 
Here, we analyze in detail the behavior of polycontextural networks close to this transition. We provide an analytical estimate of the critical point and show that the scaling properties and the space-time structure of these clusters show self-similar behavior. Our results not only contribute to a better understanding of the emergence of consensus in systems of distributed beliefs but also show that polycontextural networks are models, motivated by social systems, where susceptibility -- the sensitivity to change own beliefs -- drives the growth of consensus clusters.}

% Double-space the manuscript.

% \baselineskip24pt

% % Make the title.

\maketitle

\section{Introduction}
An abrupt change in the properties of a system upon a small change in an external condition characterizes a phase transition.
Such transitions occur in different contexts~\cite{solePhaseTransitions2011}: If water is heated above 100°C a sudden transition occurs and it starts to boil. Likewise, certain materials lose their magnetic properties, if they are heated above the Curie point~\cite{cusackElectricalMagneticProperties1958}. Besides these transitions, statistical physics also knows the concept of \textit{geometric} phase transitions, e.g. in percolation theory~\cite{staufferIntroductionPercolationTheory2018} or random-graph theory~\cite{erdosEvolutionRandomGraphs1960}: At a critical fraction of added links a network of disconnected clusters merges into a large, system-wide cluster. However, phase transitions are not limited to the physical world, and they have applications in the social and behavioral sciences as well ~\cite{levySocialPhaseTransitions2005}: They can be observed in negotiation and opinion dynamics~\cite{baronchelliNonequilibriumPhaseTransition2007, mukherjeeDisorderinducedPhaseTransition2016,mansouriPhaseTransitionSocial2020}, collective decision-making~\cite{tsarevPhaseTransitionsCollective2019} and models of social influence~\cite{castellanoNonequilibriumPhaseTransition2000}. Also, the formation of social groups bears similarities to phase transitions~\cite{schweitzerSocialNucleationGroup2022}.

An important aspect of phase transition phenomena is that the observed macroscopic property is the result of a large number of microscopic interactions. In social systems, these interactions require agents to perceive and interpret transmitted signals in the same way. However, as we know from human communication research as well as social psychology, interacting persons often interpret facts differently because they have different beliefs~\cite{millerBlissfullyHappyReady2016, edwardsThatNotWhat2017,edwardsListeningMessageInterpretation2011,BergerMiscommunicationCommunication}. 
Here, belief systems refer to a ``set of predispositions within an individual to perceive, construe, and interpret stimuli or events in a consistent manner''~\cite{belchBeliefSystemsDifferential1978}. Our individual belief system hence provides ways of ``construing or dimensionalizing relevant aspects of the world''~\cite{belchBeliefSystemsDifferential1978}.

The mechanisms that lead to consensus have been analyzed by different groups and in various models. However, most conventional models presuppose a shared understanding of facts and hence assume that perceived beliefs represent objective truth. These models hence neglect any subjectivity and possible \textit{external dissonance}~\cite{dalegeNetworksBeliefsIntegrative2023}, although their importance is backed by social influence studies. However, there are approaches to modeling certain types of subjectivity in agent-based models. Some approaches use fuzzy logic to account for the fact that human opinions are often fuzzy and that the communication of opinions can often be ambiguous or vague~\cite{wangFuzzyOpinionNetworks2016,gasparriFuzzyOpinionDynamics2012}. In these models, the shape and width of the membership functions represent the uncertainty about a specific opinion. 
%In contrast to fuzzy logic, our model does not contain any uncertainty components. Each agent bases its decision on a crisp binary logic. However, because of the different translation tables, each agent's logic is based on a different definition of facts/colors.
    A very different form of subjective perception is analyzed in agent-based models that distinguish between observable expressed and hidden private opinions (EPOs)~\cite{yeInfluenceNetworkModel2019,houOpinionDynamicsModified2021}. Depending on whether expressed and private opinions are equal when reaching a consensus, these models allow for distinguishing between different types of conformity (e.g. conversion, compliance, or disinhibitory contagion)~\cite{jedrzejewskiThinkThenAct2018}. Additionally, EPO models enable the analysis of how perceived group pressure can influence private opinions~\cite{chengConsensusExpressedPrivate2020} or how private opinions can be inferred by observing other's behavior~\cite{dongOpinionFormationAnalysis2024}. 
%Most EPO models share with our model that the private opinion (in our model the private belief system) and the opinion observed by others can be different. However, in contrast to our model agents are aware of these differences. If present, they know about their public conformity without private acceptance. Furthermore, our model supposes that each agent works according to its own logical logic. In contrast, in EPO models opinions are understood globally equal. Consequently, the stable misunderstandings we observed are not possible in EPO models.}

To answer whether and how people with different belief systems can reach a consensus, we proposed \textit{polycontextural networks}~\cite{falkCollectivePatternsStable2022a}, a model of interacting agents that strive for consensus, however, each with only a subjective perception of the world. Depending on a parameter that governs how responsive the agents are to changing their beliefs, we observed a phase transition between a system with permanent changes of the beliefs and a frozen state where no belief changes can be observed. A build-up of aligned clusters was only observable around the transition point. In contrast to fuzzy logic, our model does not contain any uncertainty components. Each agent bases its decision on a crisp binary logic. However, because of the different translation tables, each agent's logic is based on a different definition of facts/colors. Most EPO models have in common with our model that the private opinion (in our model the private belief system) and the opinion observed by others can be different. However, in contrast to our model agents are aware of these differences. If present, they know about their public conformity without private acceptance. Furthermore, our model assumes that each agent only perceives according to its own logical logic. In contrast, in EPO models opinions are understood in the same way by every agent. Consequently, the stable misunderstandings we observe are not possible in EPO models.

In our initial publication~\cite{falkCollectivePatternsStable2022a}, we used the term \textit{worldview} instead of \textit{belief system}. The distinction between both terms in the scientific literature is rather blurred and, depending on the scientific discipline, they are interchangeably used to describe a set of beliefs that we use to describe and make sense of reality~\cite{sireNamingElephantWorldview2015, belchBeliefSystemsDifferential1978, koltko-riveraPsychologyWorldviews2004,uso-domenechWhatAreBelief2016b}. The term \textit{worldview} (in German \textit{Weltanschauung}) has been introduced by the philosopher Kant and has since also been used in other scientific disciplines like theology~\cite{naugleWorldviewHistoryConcept2002,johnsonIntegratingStudyCulture2011,vanderkooijWorldviewMeaningConcept2013}, psychology~\cite{koltko-riveraPsychologyWorldviews2004}, cultural sciences~\cite{haarmannFoundationsCulture2021} and sociology~\cite{mifsudWorldviewsRoleSocial2023}. Here, worldview usually refers to a self-contained and comprehensive framework of mutually compatible beliefs. In contrast, belief systems can refer to an interrelated subset of beliefs that, only when coherently clustered, are recognizable as generalized worldviews~\cite{mifsudWorldviewsRoleSocial2023}. While the philosophical implications of the model played a greater role in our initial publication, the concept of beliefs seems more appropriate for mathematical consideration~\cite{dalegeNetworksBeliefsIntegrative2023,rodriguezCollectiveDynamicsBelief2016b}, and thus also for this paper.

In our initial publication of \textit{polycontextural networks}, we focused on the philosophical foundation and sociological motivation. We showed that the model serves as an implementation of the theories of the philosopher Gotthard Günther~\cite{guentherBeitraegeZurGrundlegung1976, falkPhysicsOrganizedTransformations2021}. As such, it might enable the analysis of the emergence of self-consciousness in the interaction of subjectively perceiving agents. We showed how different agents cannot synchronize if they adapt their belief system too slowly or too quickly. The presented results were mainly on a phenomenological level. However, the sudden change in the size of the belief-aligned cluster shows characteristics of what is known in statistical physics as a phase transition. Therefore, in this paper we apply methods from statistical physics to analyze the transition phenomenon observed in our social model, as is characteristic of the field of social physics~\cite{jusupSocialPhysics2022,pentlandSocialPhysicsHow2015}.
We analyze the space-time structure, as well as the scaling properties that appear in our polycontextural network model. We find that our model shows scale invariance in the spatial as well as in the time domain. Hence, similar to what is known for opinion dynamics models~\cite{toralFiniteSizeEffects2006,joFinitesizeEffectsConvergence2021}, our results show the non-trivial correlations between the dynamics of our model, the convergence time and the system size.
Additionally, we propose a mechanistic explanation of our model's dynamics close to the critical point. 
Since polycontextural networks serve as a generic model for the build-up of shared beliefs, our detailed analysis of the behavior close to the phase transition helps to better understand phenomena like opinion polarization and the stabilization of opinion communities.

The remainder of the manuscript is as follows. In the next section, we recall the definition as well as the basic characteristics of the polycontextural networks model and introduce the threshold parameter $q$. In section~\ref{sec:analytic} we derive analytical estimates for the upper and lower bounds of $q$. We then analyze the scaling behavior and the space-time structure of belief-aligned clusters in~\ref{sec:scaling}, before we summarize our results in the last section.

\section{Model}
\label{sec:model}
\begin{figure*}
        \centering
        \includegraphics[width=1\linewidth]{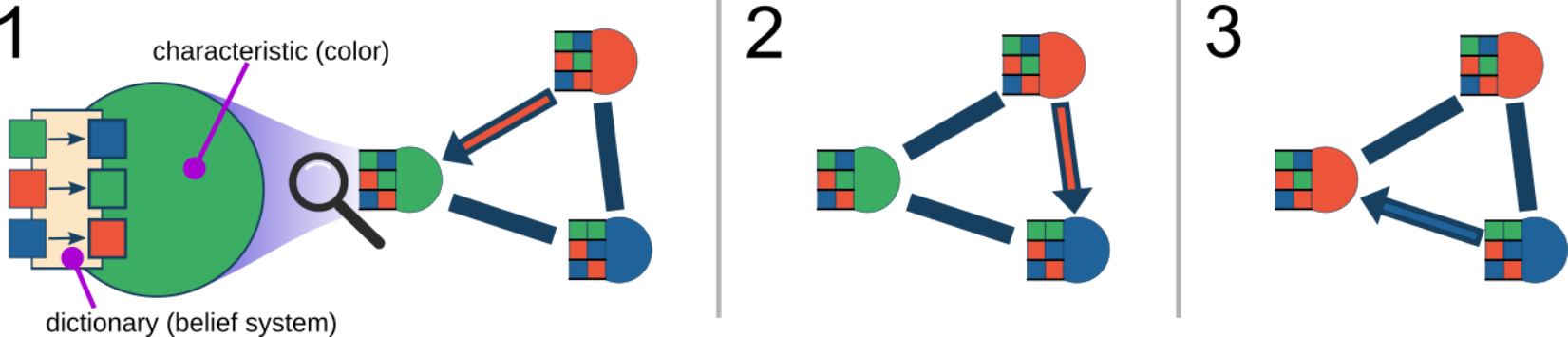}
        \caption{Illustration of the model's characteristics and its dynamics. 1: Each agent is equipped with a dictionary that maps the outside world to its internal perception. E.g. something that is \textit{green} from an observer's perspective is perceived by the shown agent as \textit{blue}. Additionally, each agent has a characteristic, a color. In the shown interaction of three agents, the agent with green color observes the characteristic of the orange agent. Since the green agent's dictionary maps orange to green, the agent perceives green and doesn't change its color. 2: The blue agent observes the orange agent's characteristic. Since the blue agent's dictionary maps orange to blue, it also does not change its color. 3: The initial green agent observes the blue agent. The green agent's dictionary maps blue to orange, hence the agent changes its color from green to orange. }
        \label{fig:dynamics}
    \end{figure*}

The definition of the model is equivalent to the polycontextural networks model we recently proposed elsewhere~\cite{falkCollectivePatternsStable2022a}, and is repeated here for the sake of completeness (see also Fig.~\ref{fig:dynamics}).
The polycontextural network is a stochastic model, where $N$ agents interact in a network. Each agent $A_n$ with $n \in \{1\dots N\}$ is endowed with a certain characteristic $c_n \in C$, where $C$ contains all alternative states the characteristic may have. For all agents, the set of expressions is the same. Thus, the characteristic of each agent is given as a standard basis vector $\boldsymbol{e}_i$ of length $C$ with $1$ in the $i$th position and 0 in every other position.

Additionally, each agent has an individual dictionary (its belief system) that bijectively maps the `objective outside world' of the agent to its personal cognition. Formally, this dictionary is a bijective function $\sigma: C \to C$ and can be written as a $C \times C$ permutation matrix $T_n$. If one agent $A_n$ observes the characteristic $c_m$ of another agent, the observing agent perceives $T_n c_m$ instead of the `true' (objective) $c_m$. The result of such an observation can be interpreted as the agent's belief about the other agent's actual belief~\cite{dalegeNetworksBeliefsIntegrative2023}. In the following, we will assume that the characteristics $c_n$ are colors. Due to this definition, our model does not have objective truth values -- a predefined understanding of color -- but $C!$ different and equally correct belief systems (here: color mappings). Initially, for each agent $A_n$, a belief system $T_n$ is randomly selected from the available $C!$. In contrast to most other models of belief dynamics we hence only have perceived (subjective) and no `true' (objective) beliefs.

Each agent is equipped with two internal counters: \textit{\#O\textsubscript{n}} and \textit{\#K\textsubscript{n}}, and every update step proceeds as follows: 
\begin{itemize}
\item One agent $A_n$ is randomly selected, and one of its neighbours $A_m$ is chosen. 
\item Agent $A_n$ subjectively observes the characteristic of $A_m$, denoted as $c_m$ but subjectively perceived as $T_n c_m$. 
\item If $A_n$ already has the same characteristic as the one perceived from $A_m$, only its internal counter \textit{\#O\textsubscript{n}} is increased by 1.
\item Otherwise, $A_n$ changes its own characteristic to the subjectively observed one and increments both its internal counters \textit{\#O\textsubscript{n}} and \textit{\#K\textsubscript{n}} by one.
\item If the fraction of times, represented by $\textit{\#K\textsubscript{n}}/\textit{\#O\textsubscript{n}}$, that $A_n$ has changed its characteristic exceeds the parameter $q$, the agent will randomly select a new belief system from the $C!$ available ones and reset both counters to zero. Note that \textit{\#O\textsubscript{n}} is always incremented at least once before its final step, so the fraction $\textit{\#K\textsubscript{n}}/\textit{\#O\textsubscript{n}}$ is always defined.
\end{itemize}
A sequence of $N$ updates forms one time step, which means that on average at every time step every node is selected once (one Monte-Carlo-Step).

Among the existing models, perhaps most similar to our polycontextural networks is the well-known Axelrod model~\citep{axelrodDisseminationCultureModel1997a}. In the Axelrod model, every agent is equipped with a vector of features. The more similar the features between two agents are, the higher the probability they interact and harmonize parts of their features. The model converges to a state without any further dynamics, where either all features between neighboring agents match (nothing more can be matched) or no features match (there is no more interaction). In our model, every agent is equipped with a single feature and a translation table. And here is the main difference to the Axelrod model: Due to the different belief systems (the different translation tables), there is no measure for similarity in our model, as required in the Axelrod model. Accordingly, the probability of interaction is constant and the individual agents can only learn about the differences in features through mutual interaction. Over time, the belief systems converge. However, this is not the result of predefined harmonization, but it emerges from the interactions and their consequences.

In what follows, we are mainly interested in the dynamics and the organization of the belief system (the ``color translation tables''). In this context, the characteristics of the agents are just signals to transmit information about their own belief system and will not be analyzed further. 
%For all simulations, we set and—to avoid boundary effects—use periodic boundary conditions if applicable.

In the initial publication, we showed that -- depending on the value of $q$ -- the polycontextural networks show a transition that mediates between an active phase where the belief systems of the agents change frequently, and an inactive (frozen) phase, where almost no belief system changes occur. In both phases, the belief systems are not correlated over long scales. In contrast, close to the transition point we observe long-range correlations as clusters of mutually compatible belief systems. One should note that compatible does not necessarily mean that the belief systems of two nodes $n$ and $m$ are equal, but only that their color mappings mutually agree in all colors, which means $T_n \times T_m = \mathbb{I}$.

\begin{figure*}
\includegraphics[width=\textwidth]{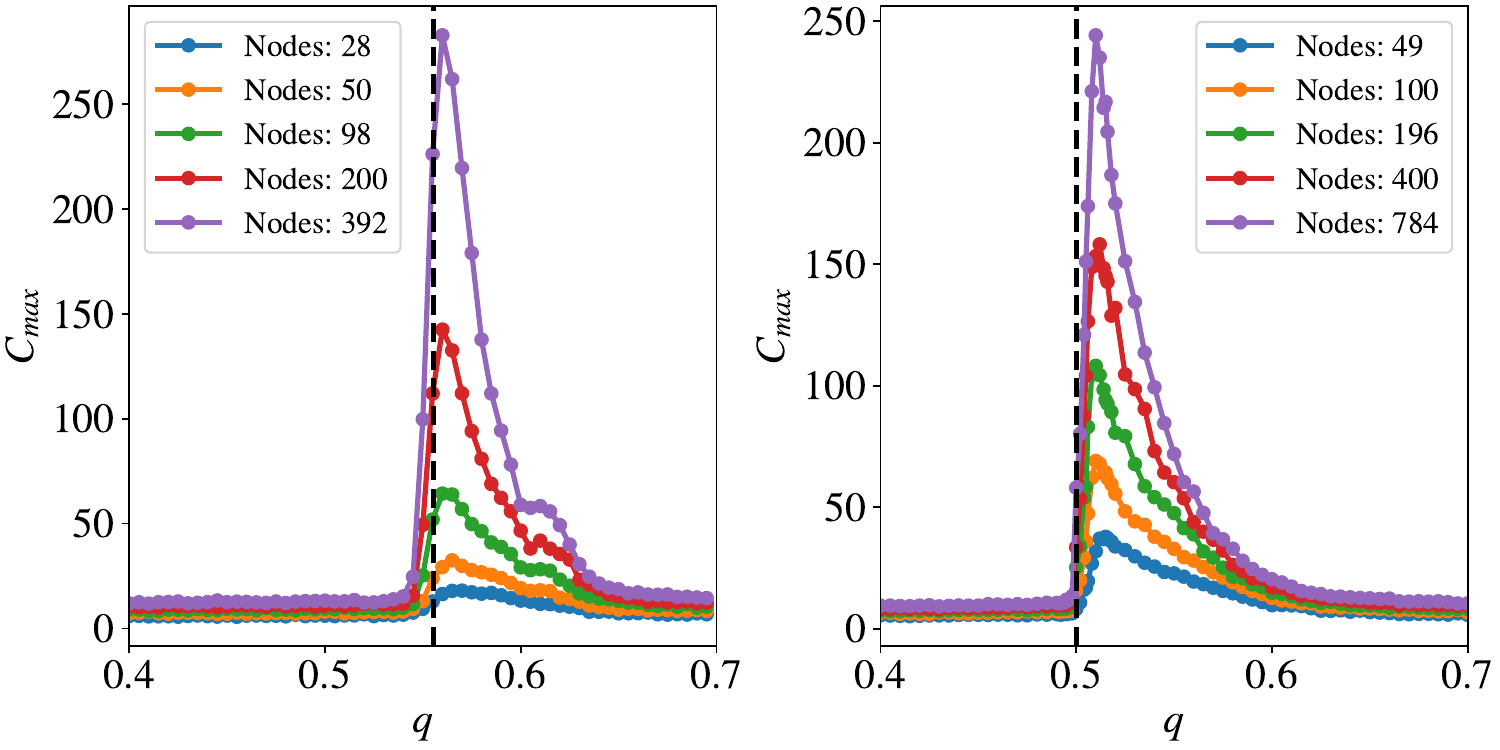}
\caption{
Size of the largest cluster ($C_{max}$) vs. the threshold parameter $q$ for (left) triangular and (right) square lattices with different numbers of nodes and $C=3$ available colors. Simulated steps: $t=20,000$, averaged over 100 runs.  the black dashed vertical lines represent the lower bound $q_-$ obtained via Eq.~\ref{eq:q_lower}. }
\label{fig:triang_grid_q}
\end{figure*}

As an example, Fig.~\ref{fig:triang_grid_q} shows the largest cluster of compatible belief systems ($C_{max}$) vs. the threshold parameter $q$ for triangular and square lattices with different numbers of nodes and after $t = 20,000$ simulated steps. As outlined above, we observe two phases $q \ll q_c$ and $q \gg q_c$, where no clusters of compatible belief systems emerge. Only for an intermediate value $q \approx q_c$ clusters of compatible beliefs can build up.

\begin{figure*}
\centering
\includegraphics[width=\textwidth]{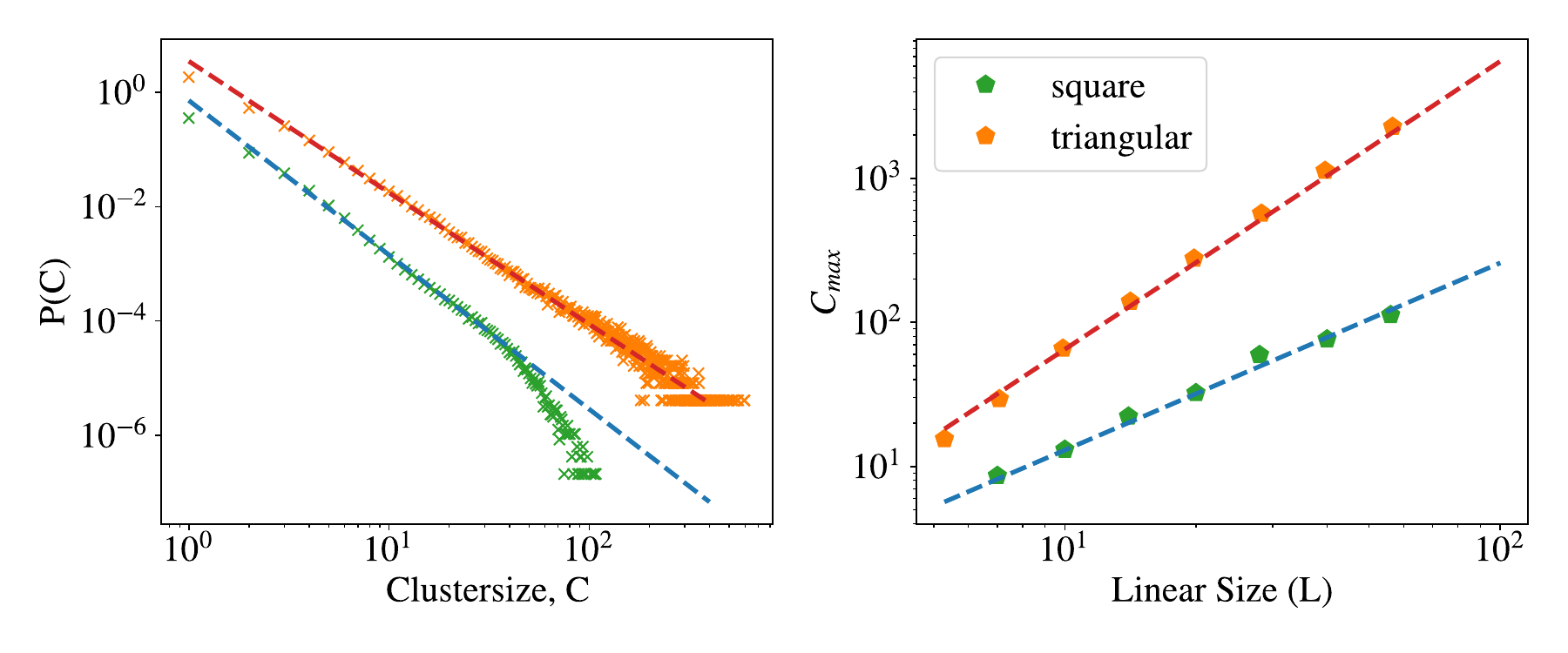}
\caption{
Left:~Cluster size distribution of the triangular network for $q=0.57$ at $t=5000$ ($N$=2450 nodes, averaged over $1000$ runs, 3 availabile colors) and of the grid network for $q=0.509$ at $t=5000$ ($N$=2500 nodes, averaged over $1000$ runs). The red line indicates a power law with exponent $\alpha = -2.3$ and the blue line indicates a power law with exponent $\alpha = -2.7$. For better visualization, we shifted the orange data in y direction.\\
    Right:~Size of the largest cluster $C_{max}$ over the linear size $L = N^{0.5}$ of the system. Both systems were simulated at $q \approx q_c$ for $t=20,000$ time steps and with 3 available colors. The dashed lines indicate the respective power law fits with (red) $d_f = 2$ and (blue) $d_f = 1.3$. (Figure partly adapted from~\cite{falkCollectivePatternsStable2022a})
}
\label{fig:cluster_dimension}
\end{figure*}

An analysis of the cluster sizes of the triangular network close to the critical point ($q \approx q_c$) revealed that during the build-up of the clusters, the distribution follows a power-law with an exponent of $\alpha = -2.3$ (Fig.~\ref{fig:cluster_dimension}~(left)). 

This means that the functional relationship can be described by:
\begin{equation}
    f(x) = c x^{\alpha},
\end{equation}
where $c$ is a constant.

Additionally, it was observed that for both network types the size of the largest cluster $C_{max}$ scales with the linear system size $L$ according to
\begin{equation}
    C_{\max}(L) \sim L^{d_f},
\end{equation}
where $d_f$ is the (possibly) fractal dimension of the cluster~\cite{lesneScaleInvariance2012,tsakirisPercolationRandomlyDistributed2010} (Fig.~\ref{fig:cluster_dimension}(right)).

In the following, we will provide a detailed analysis of the transition point, as well as the scaling properties and the space-time structure of the clusters.

\section{Separation of three different phases}
\label{sec:analytic}
As suggested by Fig.~\ref{fig:triang_grid_q}, depending on the value of the parameter $q$ our model shows three different dynamical phases. In Ref.~\cite{falkCollectivePatternsStable2022a} we provided a simple mechanistic explanation for these three phases. Here, we extend this argument and analytically derive approximate values for the boundaries between these domains. 

For our approximation, we assume random-$d$-regular networks. We consider two connected actors: $A$ and $B$. 
Let us assume that there is no correlation between the two actors, which means that at each time step $B$ randomly changes its color to one of the $C$ available. If $A$ observes the color of $B$, it perceives~$T_A c_B$.
Since the translation tables are bijective mappings, $A$'s observed color will be similar to its own color in $1/C$ cases. Neglecting stochastic fluctuations we hence know that $A_n$ already has the same characteristic as the observed one with a probability of $1/C$. Now, if $q$ is larger than $1 - 1/C$, this small fraction of correct observations is enough for $A$ to not change its translation table. The same holds for all other agents. Above the threshold of $q_{+} = 1 - 1/C$ we hence have a \textit{frozen dynamic}. Due to the limited number of colors, it is possible that small clusters exist just by accident, but the individual actors do not actively change their belief systems in order to align with neighbors.

We now turn to the lower threshold. We consider an actor $A$ that has only neighbors with the same belief system as $A$. We call the randomly selected neighbor that $A$ observes $B$. We further assume that since $A$'s last update step also $B$ has performed one update, one observation. During $B$'s last update, it observed $A$'s color with a probability of $1/d$ ($d$ is the degree of the random-regular graph). Since $A$ and $B$ share the same belief system, in this case, $B$ changed its color such that $A$ now observes the color that matches its own. With a probability of $1 - 1/d$, $B$ did not observe $A$ but another of its neighbors. Since we assumed an uncorrelated environment, with a probability of $1/C$ this color nonetheless matches the one, which $A$ wants to observe. Hence, given that $A$ and $B$ share the same belief system, the probability that $A$ observes the correct color is given by:
\begin{align}
\label{eq:lower}
   p_{-} =  \frac{1}{d} +  \frac{d-1}{d} \cdot \frac{1}{C}.
\end{align}
In the above derivation, we assumed that since $A$'s last update step also $B$ has performed one update. It might also happen that this is wrong and instead, $B$ has not performed any update. In this case, however, one can apply the same arguments to $A$ instead of $B$. The result does not change. 

If we again neglect any stochastic fluctuations, we can state that nodes requiring more than $p_{-}$ correct observations cannot stabilize, even if they share the same belief system with their neighborhood. Below the threshold of 
\begin{align}
    \label{eq:q_lower}
q_{-} = 1 - p_{-}
\end{align}
we hence have random dynamics where no clusters of shared tables can emerge.

We have shown that below $q_{-}$ no cluster can build up, since the actors require more agreement than possible even in a system with equal belief systems. Likewise, we have shown that no cluster can emerge above $q_{+}$ because the actors require less agreement than randomly observed. Within the boundaries of $q_{-}$ and $q_{+}$ the growth of clusters is possible and they are statistically stable. There is, however, a drastic difference between both limits: If $q$ is smaller, but close to $q_{+}$, small clusters consisting of only a few nodes can already form stable structures. In contrast, if $q$ is only slightly larger than $q_{-}$ the only stable cluster is the giant cluster that covers the full system. At $q_{-}$ we hence observe a sharp transition, whereas at $q_{+}$ there is a smooth crossover into the frozen regime. In Fig. 2, the size of the largest clusters near $q_-$ is smaller than the system size. The reason for this is that the dynamics of the model slow down with the simulation time and the associated few remaining large clusters. So, although the giant cluster is the only stable one, it is only reached after an extremely long simulation time.

\begin{figure*}
\includegraphics[width=\textwidth]{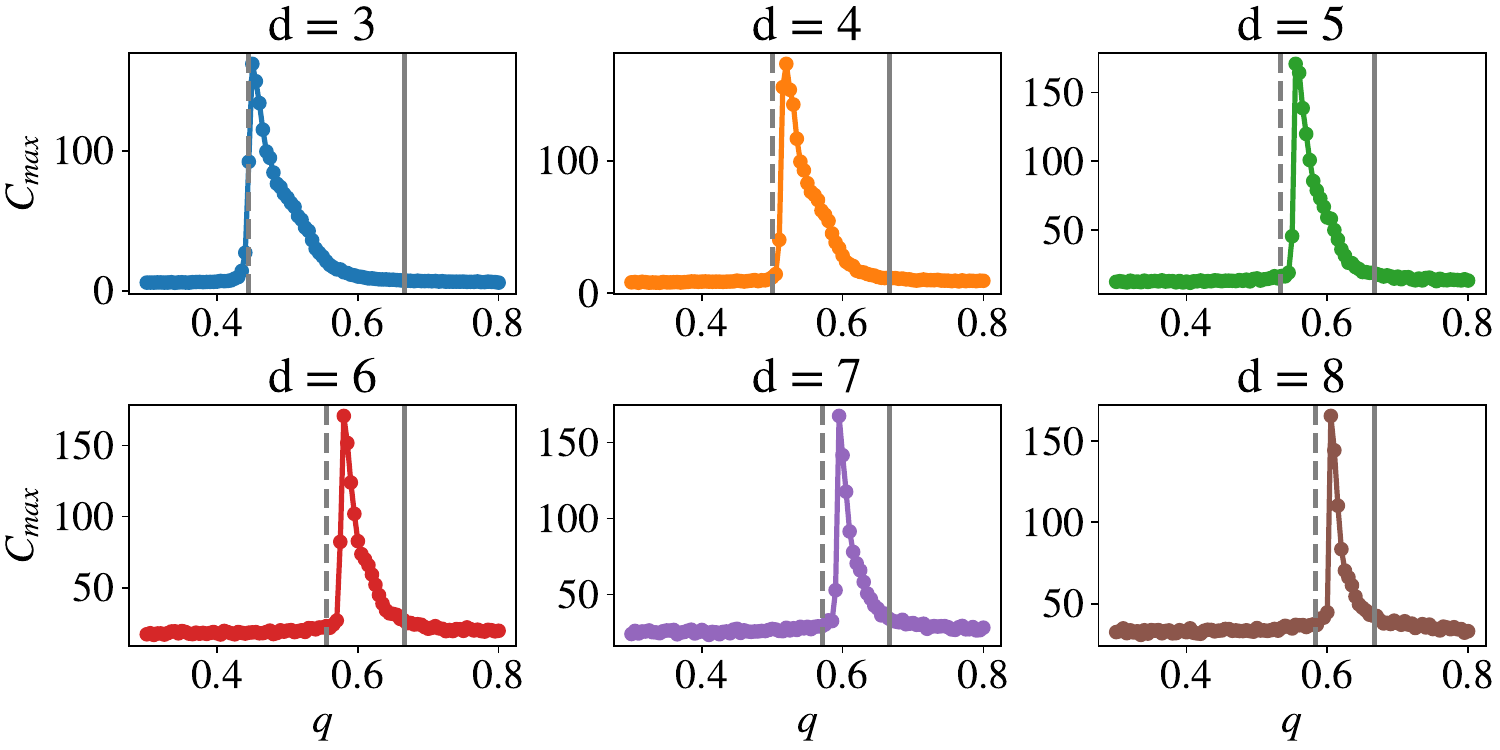}
\caption{
Size of the largest cluster ($C_{max}$) vs. the threshold parameter $q$ for random-regular networks with different degrees $d$ and 3 available colors. Simulated steps: $t=80,000$, averaged over 100 runs, N=200 nodes. The dashed line denotes the respective lower critical value ($q_{-}$), the solid line indicates the upper bound $q_{+}$.}
\label{fig:rr_q}
\end{figure*}

In Fig.~\ref{fig:rr_q} we show the size of the largest cluster vs. $q$ for random regular networks with different degrees. Additionally, we added two gray lines that denote the lower (dashed) and upper (solid) threshold for the build-up of clusters. Even though our approximation is only of the first order and neglects the possibility of structural effects (e.g. triangles in the network), we observe a strong agreement between the analytical results and the simulations.

\begin{figure*}
\includegraphics[width=\textwidth]{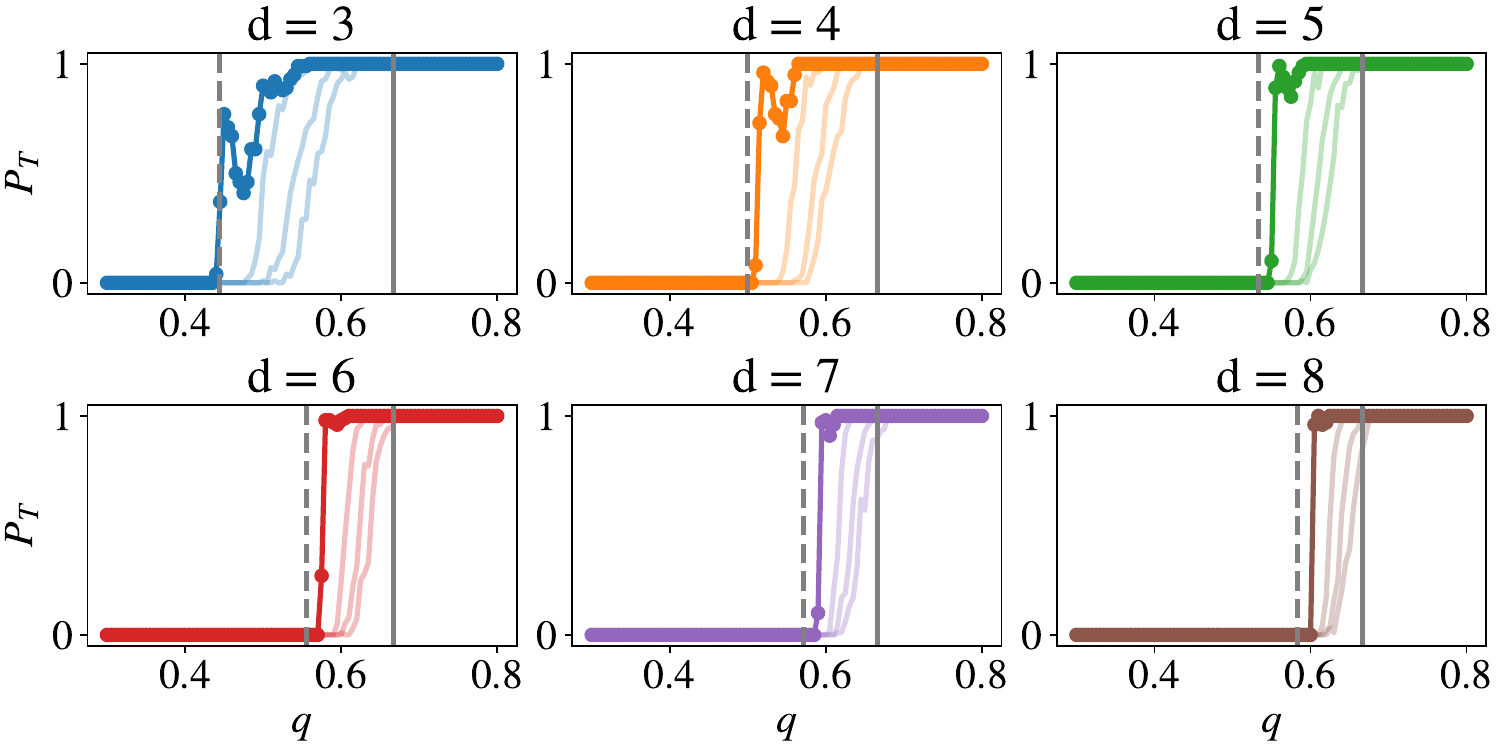}
\caption{
Order parameter $P_T$ vs. the threshold parameter $q$ for random-regular networks with different degrees $d$ and 3 available colors at different times $t=\{500,1.000,5.000,20.000\}$ (averaged over 100 runs, N=200 nodes). The saturated lines refer to $t=20.000$, and the transparent lines refer to the other times. The dashed line denotes the respective lower critical value ($q_{-}$), the solid line indicates the upper bound $q_{+}$. }
\label{fig:order_rr_q}
\end{figure*}

The described behavior is characteristic of a nucleation transition~\cite{teranTimeevolutionGrainSize2010,merNucleationPhaseTransitions1952,corberiMultinucleationFirstorderPhase2019a}, where for given parameters a nucleus starts to grow to form a large cluster that can eventually reach the size of the full system. Phase transitions like the observed nucleation transition are defined by an abrupt change in the characteristic of a system upon a small change of a macroscopic parameter, often called a control parameter. In the polycontextural networks, the control parameter is given by $q$ which governs how responsive the agents are to changing their belief system. The specific characteristic that reacts to a change of $q$ is then the number of table changes per time step. Based on this motivation, we introduce the quantity $P_T$, which measures the probability that the system has reached a frozen state (no color changes in the last time step) and serves as an order parameter of the system, i.e., as an indicator of the collective state. In Fig.~\ref{fig:order_rr_q} we show the sudden jump of the order parameter $P_T$ for random regular networks with different degrees and at different times $t=\{500,1.000,5.000,20.000\}$. The dashed line denotes the transition point as predicted by Eq.~\ref{eq:lower}. The remaining fluctuations especially apparent for networks with small degrees are due to competition between two or more large clusters. 

An important quantity when analyzing phase transitions is the susceptibility~\cite{radicchiPredictingPercolationThresholds2015}. The susceptibility indicates how sensitive the order parameter reacts to changes in the control parameter. Therefore, it peaks at the critical point where the order parameter $P_T$ has high fluctuations~\cite{thayer-baconNurturingRelationalEpistemology1997}. We have already discussed that in our system clusters cannot grow if there are either no table changes $q > q_{+}$ or if there are too many table changes $ q < q_{-}$. This corresponds to the two phases where either $P_T = 1$ or $P_T = 0$. We can therefore directly link the ability of clusters to form to the magnitude of the susceptibility. The observed size of the clusters is hence an indication of the susceptibility of the system.

For the presented approximation, we assumed random $d$-regular graphs. In what follows, we present a scaling analysis of the polycontextural network model. To make our results comparable to other simple and well-studied models that show phase transitions, we will focus on planar network topologies like triangular and square lattices with periodic boundary conditions. Since these networks have a constant degree, we can use our approximation to predict the transition point of these networks. In Fig.~\ref{fig:grid_triang_combined} (e) and (f) we show the plots for the order parameter $P_T$ for the triangular and square lattices. In our approximation, we neglected any structural effects. However, in Ref~\cite{falkCollectivePatternsStable2022a} we showed that especially triangular structures have a stabilizing effect on clusters. This explains why the transition point of the triangular network is shifted to the left compared to our approximation.

\section{Scaling}
\label{sec:scaling}

Scaling analyses are mostly performed when analysing second-order (continuous) phase transitions. However, it is also known that grain growth and particle coarsening can create statistical self-similarity, according to which the system shows self-similarity and hence scale invariance in a statistical sense~\cite{mullinsStatisticalSelfSimilarity1986}. Given that our polycontextural networks serve as a minimal model to study the build-up of believe-aligned clusters it is useful to clarify what scaling behavior these clusters exhibit.

\begin{figure*}
     \centering
    \includegraphics[width=\textwidth]{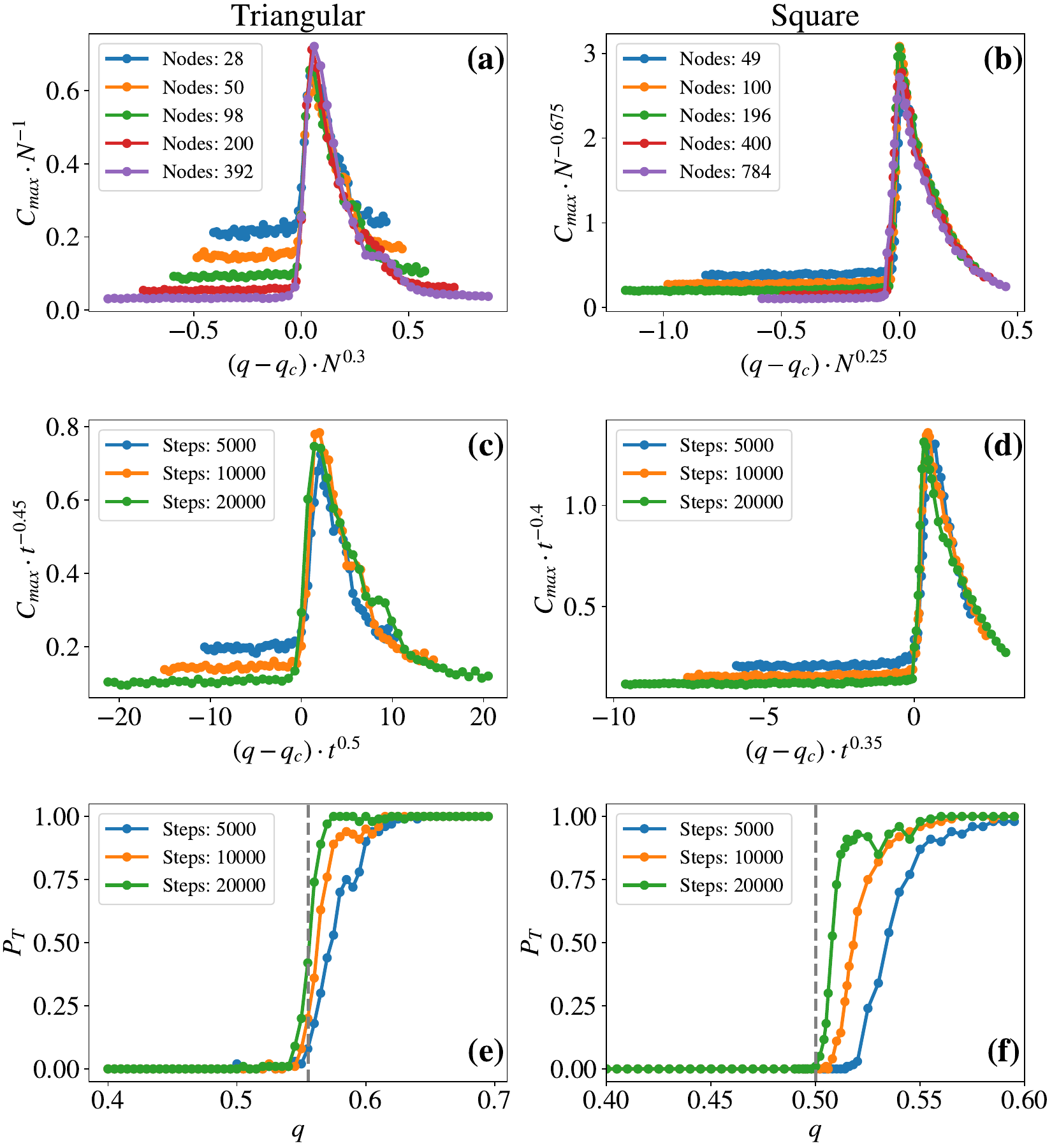}

     \caption{
     \textbf{(a)}~$C_{max} \cdot N^{-1}$ vs $(q-q_c) \cdot N^{0.3}$ for five different sizes of a triangular lattice with a simulation time of $t=20,000$ . The collapse of the curves indicates the validity of the finite-size scaling.
     \textbf{(b)}~$C_{max} \cdot N^{-0.675}$ vs $(q-q_c) \cdot N^{0.25}$ for five different sizes of a square lattice with a simulation-time of $t=20,000$. The collapse of the curves indicates the validity of the finite-size scaling.
     \textbf{(c)}~Dynamic scaling for the largest cluster in a triangular lattice. The size of the system is $N = 98$. 
     \textbf{(d)}~Dynamic scaling for the largest cluster in a square lattice. The size of the system is $N = 100$.
     \textbf{(e)}~The probability $P_T$ that the system has reached a frozen state with no colour change for a triangular lattice at different simulation times.
     \textbf{(f)}~The probability $P_T$ that the system has reached a frozen state with no colour change for a square lattice at different simulation times. The dashed line indicates the transition point as indicated by our approximation from Sec.~\ref{sec:analytic}.
     For all simulations the number of available color was set to $C=3$.
     }
     \label{fig:grid_triang_combined}
\end{figure*}

The appearance of self-similarity and scale invariance close to $q_c$ (Fig.~\ref{fig:cluster_dimension}) implies that the system (in principle) does not have a characteristic length scale $\xi_{\perp}$. In our finite system, the characteristic length is restricted by the size of the system ($L \approx \xi_{\perp}$) which means $L$ determines the relevant length scale. Contrary, for large $L/\xi_{\perp}$ the system shows an equilibrium behavior and the dynamics are almost independent of $L$ as we have already observed in Fig.~\ref{fig:triang_grid_q}. As known from the theory of finite-size scaling, the spatial~($\perp$) correlation length follows $\xi_{\perp} \sim (q-q_c)^{-\nu_{\perp}}$, which gives rise to the scaling law~\cite{tsakirisPercolationRandomlyDistributed2010}:
\begin{equation}
     P_{max}(t,L,q) = L^{-\beta/\nu_{\perp}} f((q-q_c) L^{1/\nu_{\perp}})
     \label{eq:scalel}
\end{equation}
where $f(x)$ is a scaling function and $\beta$ is an exponent that accounts for the finite simulation time and hence the limited size of the largest cluster.
In Fig.~\ref{fig:grid_triang_combined}~(a) and~(b) we apply the scaling relation to the curves for $C_{max}$, where, for better comparability between the two networks, we set $L = N$. The collapse of the curves indicates similarity in the spatial domain.

Besides the scaling with regard to the system size, our system also shows self-similarity during its time evolution. This gives rise to dynamical scaling (Family-Vicsek scaling) as it is known from cluster growth by diffusion-limited aggregation~\cite{vicsekDynamicScalingAggregation1984}:
\begin{align}
    n(t) \sim t^{-w} f((q-q_c) t^{-z}),
\end{align}
where $f(x)$ is a scaling function, and $w,z$ are dynamical exponents. The term $t^{-w}$ accounts for the decay of small clusters, and $z$ connects the cluster sizes at time $t$ with the distance from the threshold-value $q_c$~\cite{vicsekDynamicScalingAggregation1984}.
In Fig.~\ref{fig:grid_triang_combined}~(c) and~(d) we show the dynamic scaling for both the triangular and the square lattice. As expected, in both systems the curves collapse, indicating self-similarity at different times.

\section{Conclusion}

In this paper, we studied polycontextural networks for different system sizes as well as for different simulation length. We derived analytical approximations for the lower and upper bounds of the observed phase and showed their correspondence with numerical simulations. Additionally, we analyzed the scaling behavior of the belief clusters in polycontextural networks. While in a previous publication we introduced polycontextural networks as a model and described their phenomenology qualitatively~\cite{falkCollectivePatternsStable2022a}, here we focused on quantitative and analytical considerations. As known for several other models that show emerging collective behavior~\cite{castellanoStatisticalPhysicsSocial2009b}, we demonstrated how methods from statistical physics can be used to characterize the observed phenomena.

The analysis of belief systems is a current research topic in social science~\cite{millerCOVID19ConspiracyTheory2020,chaxelHowMisinformationTaints2022}. In particular, the question of how monological belief systems, closed-off clusters of mutually supportive beliefs, stabilize, is of interest~\cite{woodDeadAliveBeliefs2012}. Here, the observed dynamics are the result of a complicated interplay between one's own beliefs and the influence of others. It has recently been argued that the mutual influence of different people cannot be explained from an objective perspective. Rather, the perceived beliefs of agent $A$ depend strongly on $A$'s belief system~\cite{dalegeNetworksBeliefsIntegrative2023}. Each observation needs hence to be modeled as a subjective process. This concept of subject-based cognition has a long tradition in philosophy: Hegel's dialectic and later Gotthard Günther's theory of polycontexturality are based on this premise~\cite{guentherBeitraegeZurGrundlegung1976, falkPhysicsOrganizedTransformations2021}. With our polycontextural networks, we developed a simple system that incorporates these philosophical concepts into a numerical model. 

The phenomenon that specific views are strengthened rather than undermined by confrontation with counter-arguments is often observed in social epistemic structures like echo chambers~\cite{diazruizDisinformationEchoChambers2023}. Here, members often hold a belief system such that the existence of opposite beliefs reinforces their original belief system~\cite{nguyenECHOCHAMBERSEPISTEMIC2020}. Our model represents the same effect by assuming that the same facts are perceived and communicated differently due to different belief systems.

Thinking of the complex interaction patterns of subjects in real-world situations as a sequence of pairwise interactions is an approximation. A possible generalization of our model could hence be to go beyond these simple interactions. As a next step, it might be interesting to give the agents the ability to simultaneously observe all neighbors' colors and to infer the compatibility of the current belief system based on this information, as e.g. known from other opinion dynamics models~\cite{dongOpinionFormationAnalysis2024}.

The in-depth analysis we presented here can help to understand the emergence of belief-aligned clusters. In follow-up studies, it might then be interesting to compare the characteristics of our theoretical model with clusters of election outcomes~\cite{kimSpatialAnalysisCountylevel2003} or the structures that can e.g. be observed in interaction networks~\cite{guimeraSelfsimilarCommunityStructure2003}. A deeper awareness of the mechanisms can help to counter disinformation campaigns and opinion polarization.

\bibliography{references}

\section*{Statements and Declarations}

\textbf{Competing Interests:} On behalf of all authors, the corresponding author states that there is no conflict of interest. \\
\textbf{Author contributions:} E.E. initiated the project. J.F., M-T.H. developed the model, J.F. ran the simulations and analyzed the data. J.F. and M-T.H. wrote the manuscript. E.E., K.W. and M-T.H. supervised the project. All authors discussed the results and implications and commented on the manuscript at all stages.\\
\textbf{Data availability statement:} This is a numerical study with no experimental
data. All relevant numerical data can be reproduced with the information given in the paper.

\end{document}